\documentstyle[graphicx,epsf]{ptptex}
\title{
Numerical renormalization group study of random transverse Ising
models in one and two space dimensions}

\author{
Yu-Cheng {\sc Lin}$^{1}$,
Naoki {\sc Kawashima}$^{2}$,
Ferenc {\sc Igl\'oi}$^{3,4}$,
Heiko {\sc Rieger}$^{1,5}$
}

\inst{
$^1$NIC c/o Forschungszentrum 
    J\"ulich, 52425 J\"ulich, Germany
\\
$^2$Department of Physics, Tokyo Metropolitan University, 
    Tokyo 192-0397, Japan 
\\
$^3$Res.\ Inst.\ f.\ Solid State Physics \& Optics,   
    1525 Budapest, P.O.Box 49, Hungary
\\
$^4$Institute for Theoretical Physics,
    Szeged University, 6720 Szeged, Hungary
\\
$^5$FB 10.1 Theoret.\ Physik, Universit\"at d.\ Saarlandes,
      66041 Saarbr\"ucken, Germany
} 

\recdate{
}

\abst{The quantum critical behavior and the Griffiths-McCoy singularities of
random quantum Ising ferromagnets are studied by applying a numerical  
implementation of the  Ma-Dasgupta-Hu renormalization group scheme. We
check the procedure for the analytically tractable one-dimensional
case and apply our code to the quasi-one-dimensional double
chain. For the latter we obtain identical critical exponents as for
the simple chain implying the same universality class. Then we apply
the method to the two-dimensional case for which we get estimates for
the exponents that are compatible with a recent study in the same
spirit.}

\begin{document}

\maketitle

\newcommand{\bc}{\begin{center}}
\newcommand{\ec}{\end{center}}
\newcommand{\be}{\begin{equation}}
\newcommand{\ee}{\end{equation}}
\newcommand{\beqn}{\begin{eqnarray}}
\newcommand{\eeqn}{\end{eqnarray}}

\section{Introduction}

The effect of quenched randomness on disordered quantum magnets close
to a quantum phase transition is much stronger than  on classical
systems at temperature driven phase transitions. As first observed by
McCoy \cite{mccoy} in a somewhat disguised version of a random
transverse Ising chain, non-conventional scaling and off-critical
singularities that lead to divergent susceptibilities even away from
the critical point now appear to be a generic scenario in any
dimension, at least in disordered quantum magnets with an Ising
symmetry. The reason for this, as pointed out by Fisher \cite{fisher}
only recently, is a novel fixed point behavior of these systems under
renormalization, namely one which is totally determined by the
randomness and its geometric properties: the so called {\it infinite
randomness fixed point} \cite{fisher,motrunich}.

Within this scenario the quantum critical behavior of disordered
transverse Ising models is essentially determined by strongly coupled
clusters and their geometric properties \cite{fisher,motrunich}.  Let
$L$ be the linear size of such a cluster. Then it contributes to the
low energy spectrum with an exponentially small excitation gap of size
$\ln\Delta E \sim L^{-\psi}$, defining the exponent $\psi$. Moreover,
at the critical point, it has a total magnetization of size
$\mu\sim L^{\phi\psi}$ defining the exponent $\phi$. Finally the
linear length scale of strongly coupled clusters occurring at a
distance $\delta$ away from the critical point is
$\xi\sim|\delta|^{-\nu}$ giving rise to a third scaling exponent
$\nu$. All bulk exponents can now be expressed via $\psi,\phi$ and
$\nu$, c.f.  $\beta_b/\nu=x_b=d-\phi\psi$, $\nu_{\rm typ}=\nu(1-\psi)$
and in the Griffiths phase $z'\propto\delta^{-\nu\psi}$. For the 1d
case, as treated above, it is $\psi=1/2$, $\phi=(\sqrt{5}+1)/2$ and
$\nu=2$ for uncorrelated disorder.

The basic geometric objects, the strongly coupled clusters, still have
to be defined and this will be done within a renormalization group
scheme. However, for site or bond {\it dilution} it is immediately
obvious what these clusters are: simply the connected clusters. Hence
the critical exponents defined above are directly related to the
classical percolation exponents \cite{senthil}: Let $\delta=p-p_c$ be
the distance from the percolation threshold, $\nu_{\rm perc}$ the
exponent for the typical cluster size, $D_{\rm perc}$ the fractal
dimension of the percolating cluster, $\beta_{\rm perc}$ the exponent
for the probability to belong to the percolating cluster. Then one has
for the critical exponents defined above
\be
\nu  = \nu_{\rm perc}\;,\quad
\psi = D_{\rm perc}\;,\quad
\phi = (d-\beta_{\rm perc}/\nu_{\rm perc})/D_{\rm perc}
\ee
Next we consider the question, what happens for generic disorder
(i.e.\ no dilution, but random bonds and/or fields) and we consider
the model defined by the Hamiltonian
\be
H = -\sum_{\langle i,j\rangle} J_{ij} \sigma^z_i \sigma^z_{j} -
\sum_i h_i \sigma^x_i \;. 
\label{Hamiltonian}
\ee
Here the $\{\sigma^\alpha_i\}$ are Pauli spin matrices, and the
nearest neighbor interactions $J_{ij}$ and transverse fields $h_i$ are
both independent random variables distributed uniformly:
\beqn
\pi(J_{ij})&=&\cases{1,&for $0<J_{ij}<1$\cr
                0,&otherwise\cr}\;\nonumber\\
\rho( h_i)&=&\cases{ h_{0}^{-1},&for $0< h_i< h_0$\cr         
                0,&otherwise\cr}\;.\nonumber
\eeqn
For this case the distance $\delta$ from the critical point is
conveniently given by $\delta=\frac{1}{2}\ln h_0$. In one space
dimension this model has been investigated intensively over the recent
years \cite{fisher1d,youngrieger,big1d,ir2,oned}, and many analytical
as well as numerical tools are at hand to analyze it. Beyond the
simple one-dimensional geometry one has to rely on numerical
techniques like quantum Monte-Carlo simulations (as in the
two-dimensional case \cite{qmc}) or the numerical implementation of
the renormalization group scheme, which we outline in the next
section.

\section{The renormalization-group scheme}

The strategy of the renormalization-group \`a la Ma, Dasgupta and Hu
\cite{dasgupta} is to decrease the number of degrees of freedom and
reduce the energy scale by performing successive decimation
transformation in which the largest element of the set of random
variables $ \{h_{i;\, }J_{ij}\} $ at each energy scale is eliminated
and weaker effective couplings are generated by perturbation theory.

\begin{figure}[ht]
 \label{1ds}
 \centering
 \includegraphics[width=8cm]{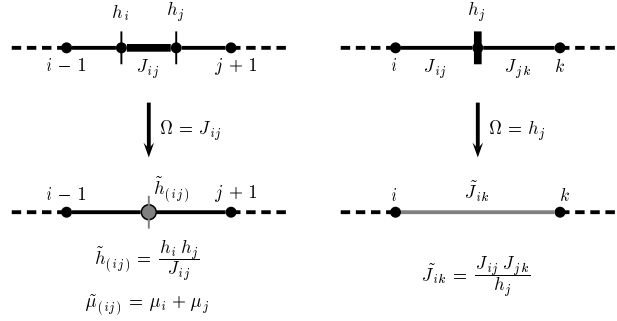}
 \caption{Schematic of renormalization-group decimation for spin chains}
\end{figure}

The renormalization-group procedure is as follows: Find the strongest coupling
\[\Omega \equiv \max \{J_{ij},\, h_{i}\}\]
in the system.  If $ \Omega =J_{ij} $, then the neighboring transverse
fields $h_i$ and $h_j$ can be treated as a perturbation to the term $
-J_{ij}\sigma _{i}^{z}\sigma _{j}^{z} $ in the Hamiltonian
(\ref{Hamiltonian}); The two spins involved are joined together into a
spin cluster with an effective transverse field
\[\widetilde{h}_{(ij)}\approx \frac{h_{i}h_{j}}{J_{ij}}\]
and an effective magnetic moment
\[\widetilde{\mu }_{(ij)}=\mu _{i}+\mu _{j} \; .\]
The bonds of the new cluster
$ \widetilde{\sigma }_{(ij)} $
with other clusters
$ \sigma _{k} $
are
\[\widetilde{J}_{(ij)k}\approx \max (J_{ik},\, J_{jk}) \; .\] 
If instead $ \Omega =h_{j} $, then the associated spin $ \sigma _{j} $
is eliminated and effective bonds between each pair of its neighboring
spins are generated by second-order perturbation theory. The strength
of the effective bonds for each pair $ (i,\, k) $ is
\[\widetilde{J}_{ik}\approx \max (J_{ik}\, ,\, \frac{J_{ij}J_{jk}}{h_{j}})\]
where the $J_{ik}$ are the bonds that may have already been present.
This procedure is sketched for the 1d case in Fig.\ 1.
We continue the procedure until there is only one remaining spin cluster.

At each stage of the RG, an effective field (bond) is a ratio of a
product of some number $ f $ of original fields (bonds)to a product of
original $ f-1 $ bonds (fields). The $ f $ grows under renormalization
at criticality. As a result, the log-field and log-bond distributions
$ R_{\Omega }(\ln \widetilde{h}) $ and $ P_{\Omega }(\ln
\widetilde{J}) $ become broader and broader under renormalization 
as the critical point is approached. This increasing width of the field
and bond distributions reduces the errors made by the second-order
perturbation approximation.  The RG becomes thereby asymptotically
exact.

\section{The one-dimensional case}

\begin{figure}[!ht]
 \label{1ddis}
 \includegraphics[angle=270,width=\halftext]{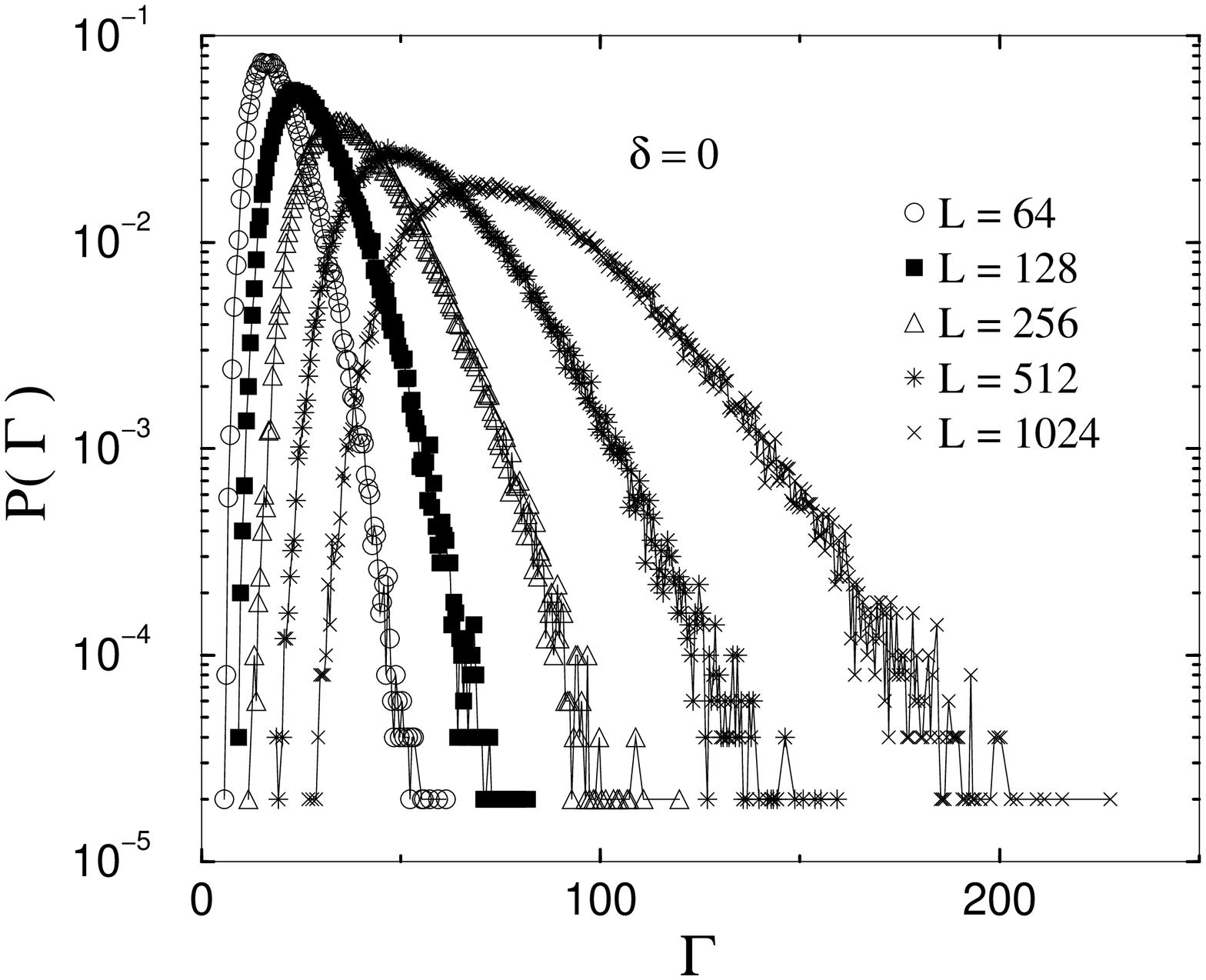}
 \hfill{} 
 \includegraphics[angle=270,width=\halftext]{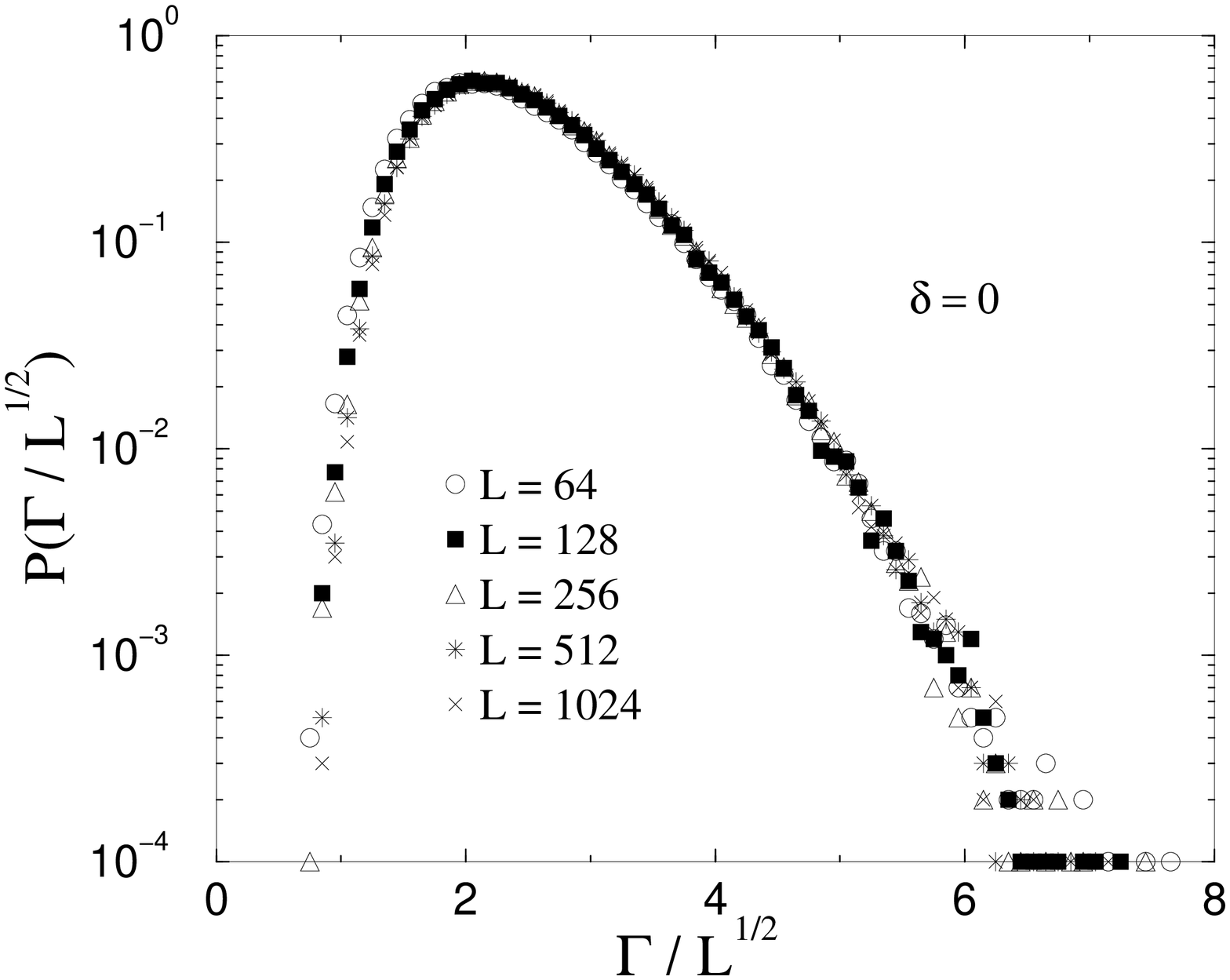}
 \caption{{\bf Left:} Distribution of the logarithmic strength of the last 
   remaining cluster fields, $\Gamma \equiv \ln
   (\frac{\Omega_{0}}{\Omega _{\tilde{h}}})$ ($\Omega_0$ denotes the energy 
   scale of the original Hamiltonian). The distribution gets
   broader on a logarithmic scale with increasing system size,
   indicating a infinite dynamical exponent $z$. The data is obtained
   from $100\, 000$ samples for each system size. 
   {\bf Right:} Scaling of the data in the left figure, assuming the 
   exponential scaling form obtained from the analytical
   work.\cite{fisher1d}}
\end{figure}

\begin{figure}[!ht]
 \centering
 \label{1dmL}
 \includegraphics[angle=270,width=\halftext]{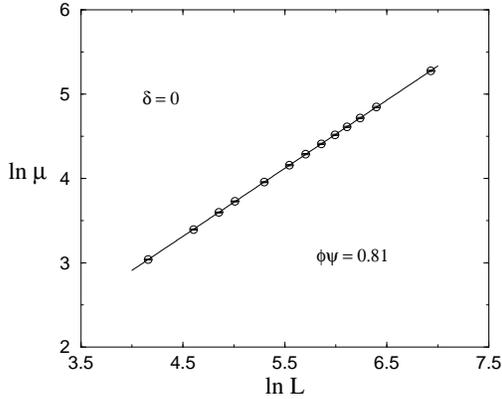}
 \begin{minipage}[t]{\halftext}
 \caption{
   Scaling of the number of active spins (proportional to average 
   magnetic moment per spin, $\mu$) in the last remaining spin cluster 
   at the critical point. We find $\mu \sim L^{0.81}$ implying 
   $\phi \approx 1.62$.}
 \end{minipage}
\end{figure}

The RG can be carried out analytically in one space dimension
\cite{fisher1d}, therefore we can use the $1d$ case with periodic 
boundary conditions as a simple check for our numerical implementation. 
In Fig.\ \ref{1ddis} we show the probability distribution of the
logarithm of the last remaining cluster field at the critical point
$h_0=1$, which scales, according to Fig.\ 2, 
like $\L^{-1/2}$, where $L$ is the system size.  From this one concludes
that the exponent $\psi$ defined in the introduction, is given by
$\psi=1/2$. Inspecting the number of active spins in the last
remaining cluster at the critical point we obtain the size dependence $\mu
\sim L^{0.81}$ from Fig.\ \ref{1dmL}, and thus $\phi \approx
1.62$.

\begin{figure}[!ht]
 \label{1dMd}
 \includegraphics[angle=270,width=\halftext]{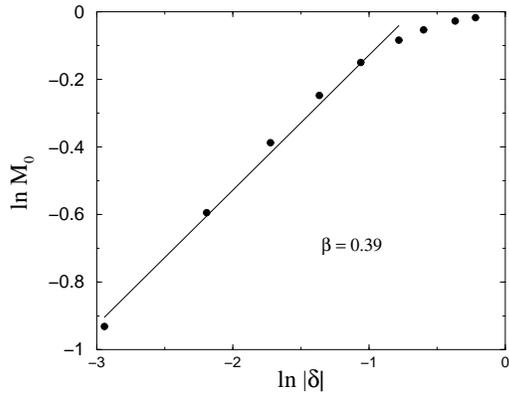}
 \begin{minipage}[t]{\halftext}
 \caption{
   In the ordered phase ($\delta < 0$), 
   the spontaneous magnetization scales as $ M_{0}=\left| \delta
   \right| ^{\beta }$ with $ \beta \approx 0.39 $.  This numerical
   estimate of $\beta $ is in agreement with the analytical
   prediction: $\beta =\frac{3-\sqrt{5}}{2}$.  Our data is obtained
   by averaging over $100\,000$ samples of size $L=1024$.}
  \end{minipage}
\end{figure}

\begin{figure}[!ht] 
  \label{1dh2}
  \includegraphics[angle=270,width=\halftext]{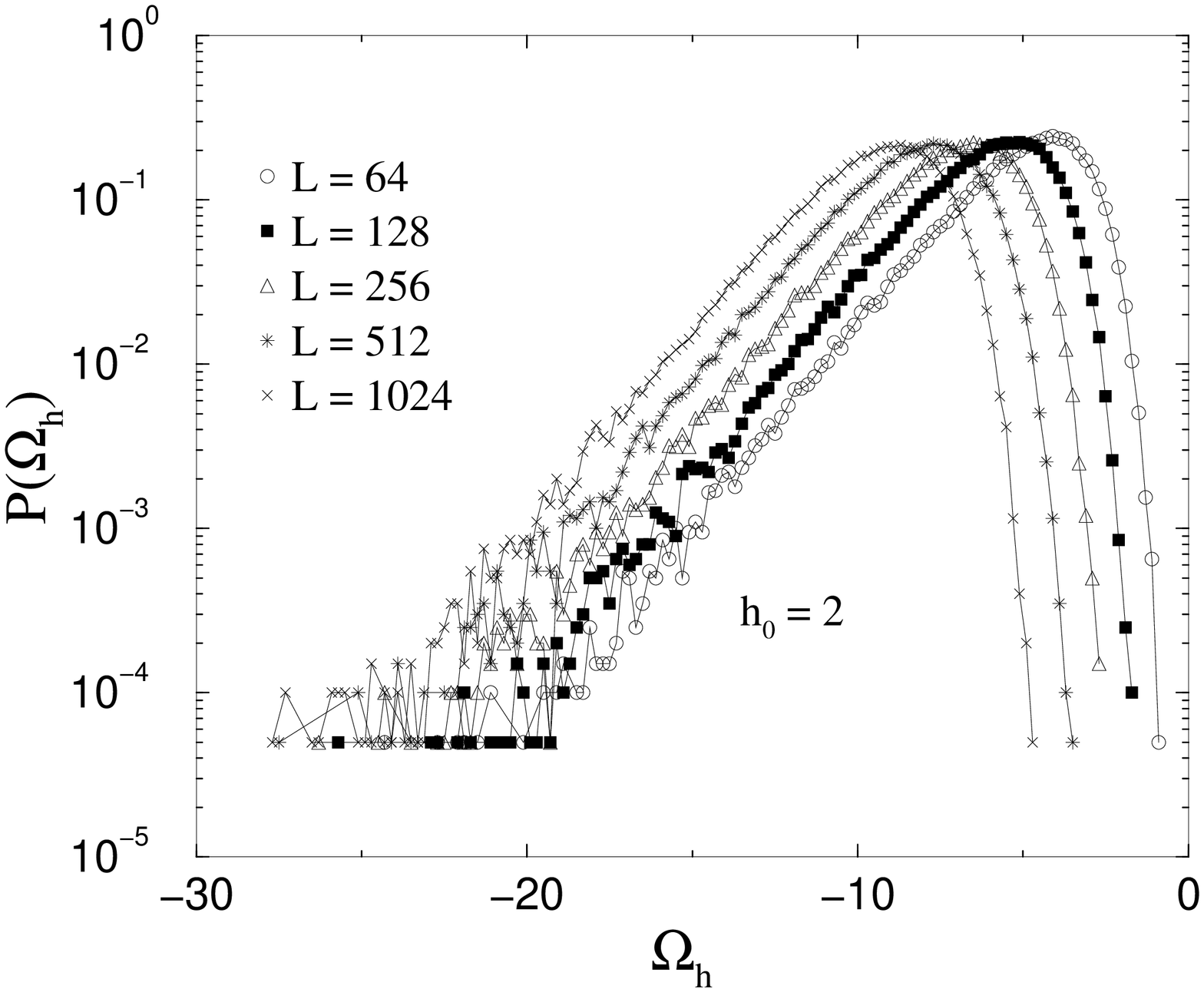}
  \hfill{}
  \includegraphics[angle=270,width=\halftext]{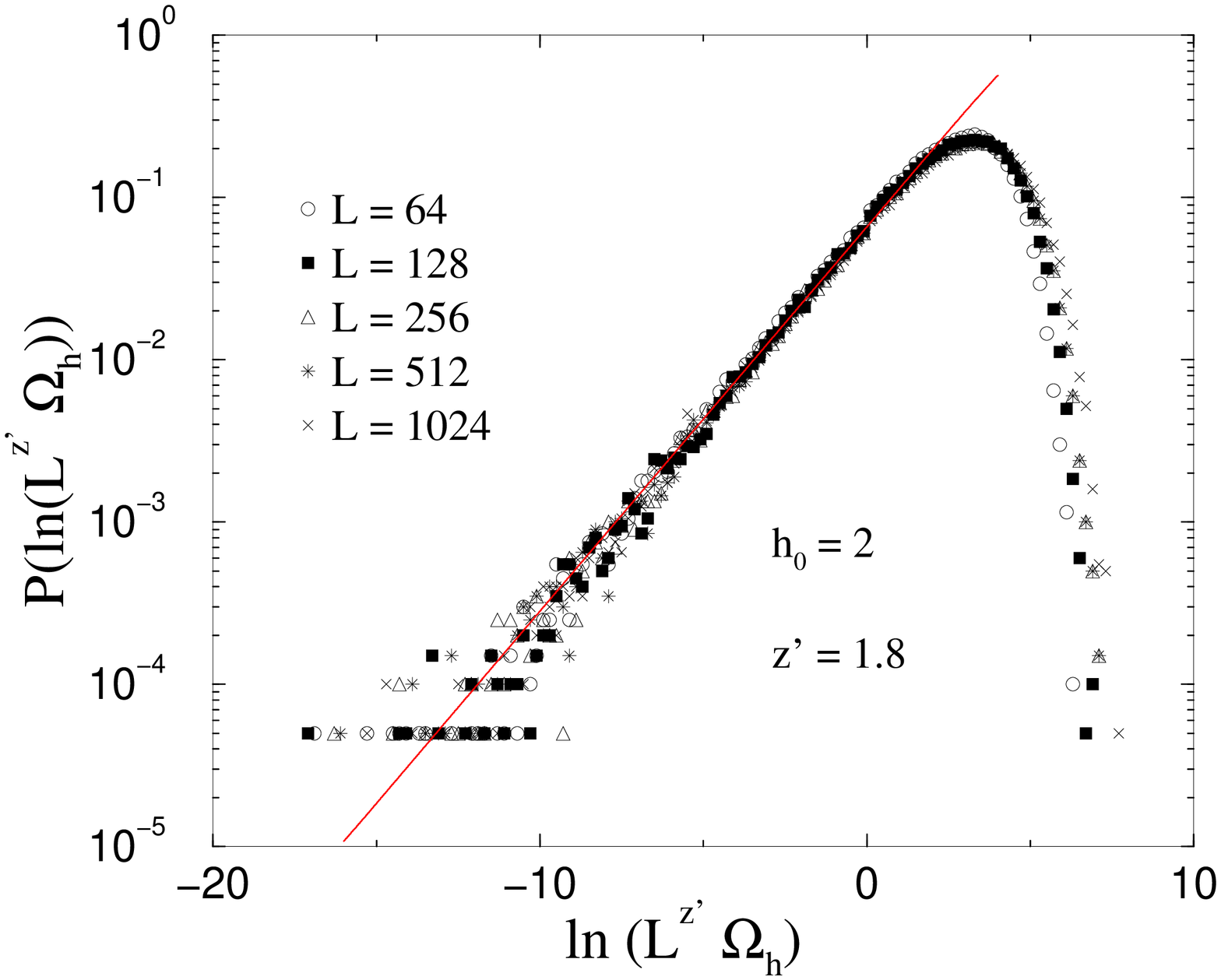}
  \caption{{\bf Left:}
    Distribution of logarithmic effective fields $\Omega_{h}$ of the
    last spin cluster in the disordered phase at $ h_{0}=2 $.  The
    curves for different sizes look very similar but shifted
    horizontally related to each other.
    {\bf Right:} Scaling plot of the data in the left figure. 
    The typical fields and spacing of rare large strongly coupled
    clusters in the disordered phase is related via $\Omega \sim L^
    {z'}$.  The dynamical exponent $ z'(h) $ is obtained from the
    asymptotic form $\ln P(\ln \Omega_{h})=1/z'(h)(\ln \Omega_{h})$+
    const.\cite{youngrieger}}
\end{figure}

\begin{figure}[!ht]
 \label{1dz}
 \includegraphics[angle=270,width=\halftext]{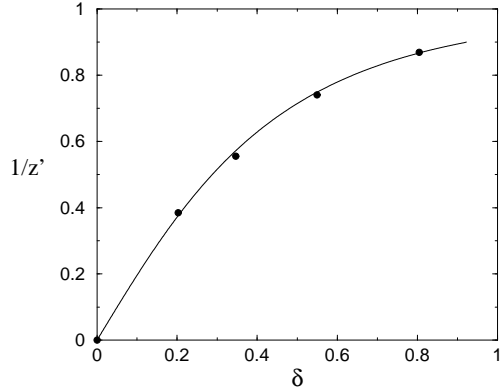}
 \begin{minipage}[t]{\halftext}
 \caption{The value of $1/z'(\delta )$ for $d=1$ obtained from the probability
   distribution of logarithmic effective fields of the last remaining
   clusters in the Griffiths-McCoy region.  The solid line, which fits
   well, is a plot of the exact $ z'(\delta ) $-relation reported in
   Ref.~\cite{ir2}: $ z'\log (1-z'^{-2})=-2h_0$.}
 \end{minipage}
\end{figure}

In the Griffiths phase $h_0\ne1$, the probability distribution of the
energy gap $\Omega$ still has an algebraic singularity at $\Omega=0$,
and its finite size scaling behavior is
\be
\Omega P_L(\Omega)\propto 
L^d\Omega^{d/z'(\delta)}
= (L^{z'(\delta)}\Omega)^{d/z'(\delta)}
\ee
where $d$ is the space dimension (in this section $d=1$) and
$z'(\delta)$ a generalized dynamical exponent that varies continuously
with the distance $\delta$ from the critical point. This exponent
parameterizes the strength of all singularities in the off-critical
region $\delta\ne0$, for instance in the disordered phase $\delta>0$
one has for the imaginary time autocorrelations $G_{\rm loc}(\tau)
=[\langle\sigma_i^x(\tau)\sigma_i(0)\rangle_{T=0}]_{\rm
  av}\sim\tau^{-1/z'(\delta)}$, for the local susceptibility
$\chi_{\rm loc}\sim T^{1/z'(\delta)-1}$, for the specific heat $C\sim
T^{1/z'(\delta)}$ and for the magnetization in a longitudinal field
$M\sim H^{1/z'(\delta)}$. The most convenient way to determine this
exponent is, however, via the distribution $P_L(\Omega)$.  At the
critical point this distribution has to merge with the critical
distribution discussed above --- and therefore $\lim_{\delta\to0}
z'(\delta)=\infty$. Using this  finite-size scaling form for the
distribution of the last bonds/fields in the RG procedure we can
extract the dynamical exponent as is done in Fig. 5.

In the ordered phase $h_0<1$ the distribution of fields and bonds 
are related to the distribution in the disordered phase $h_0>1$ via
duality, see Fig.\ 7.

\begin{figure}[!ht]
 \label{h05}
 \includegraphics[angle=270,width=\halftext]{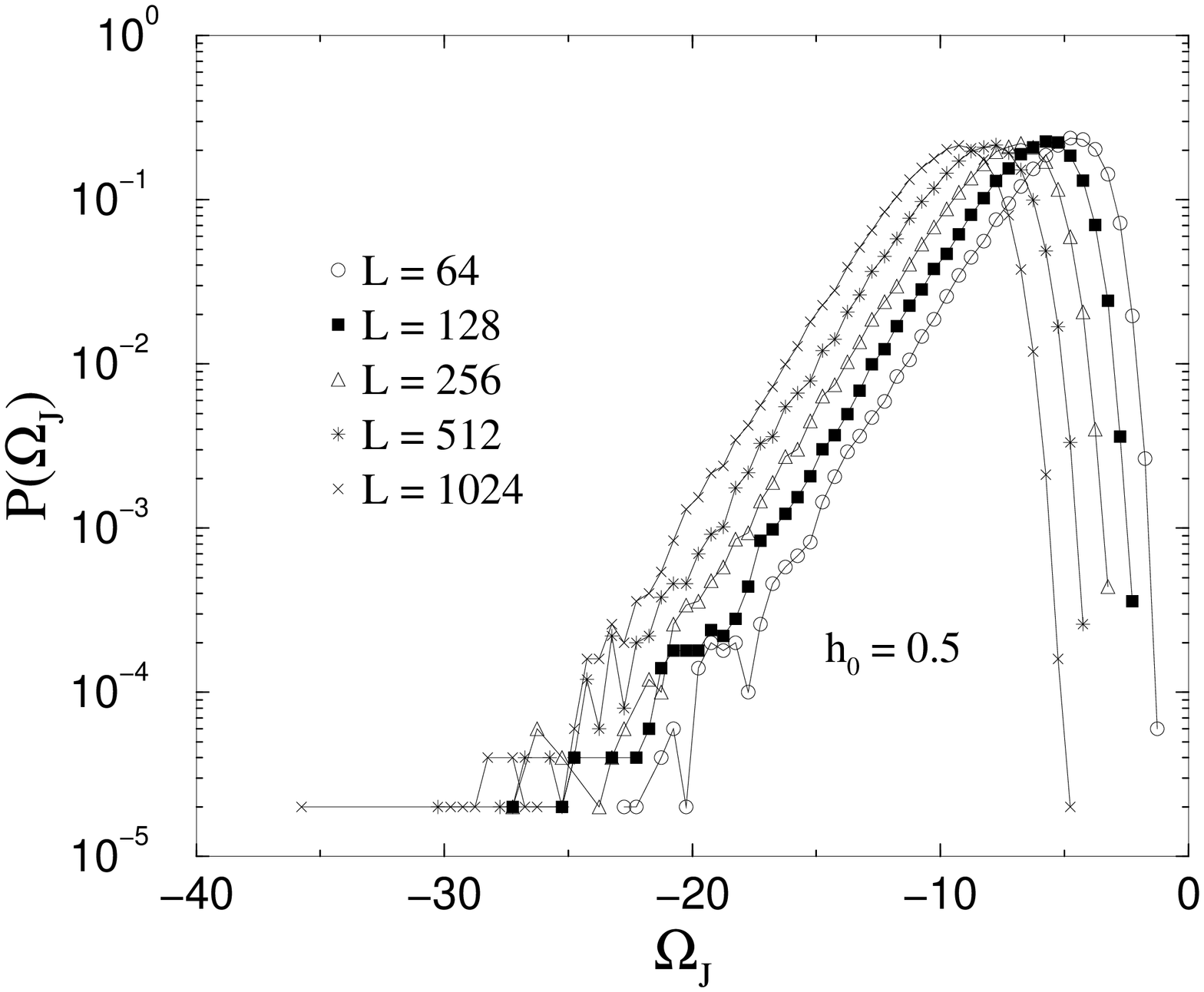}
 \hfill{}
 \includegraphics[angle=270,width=\halftext]{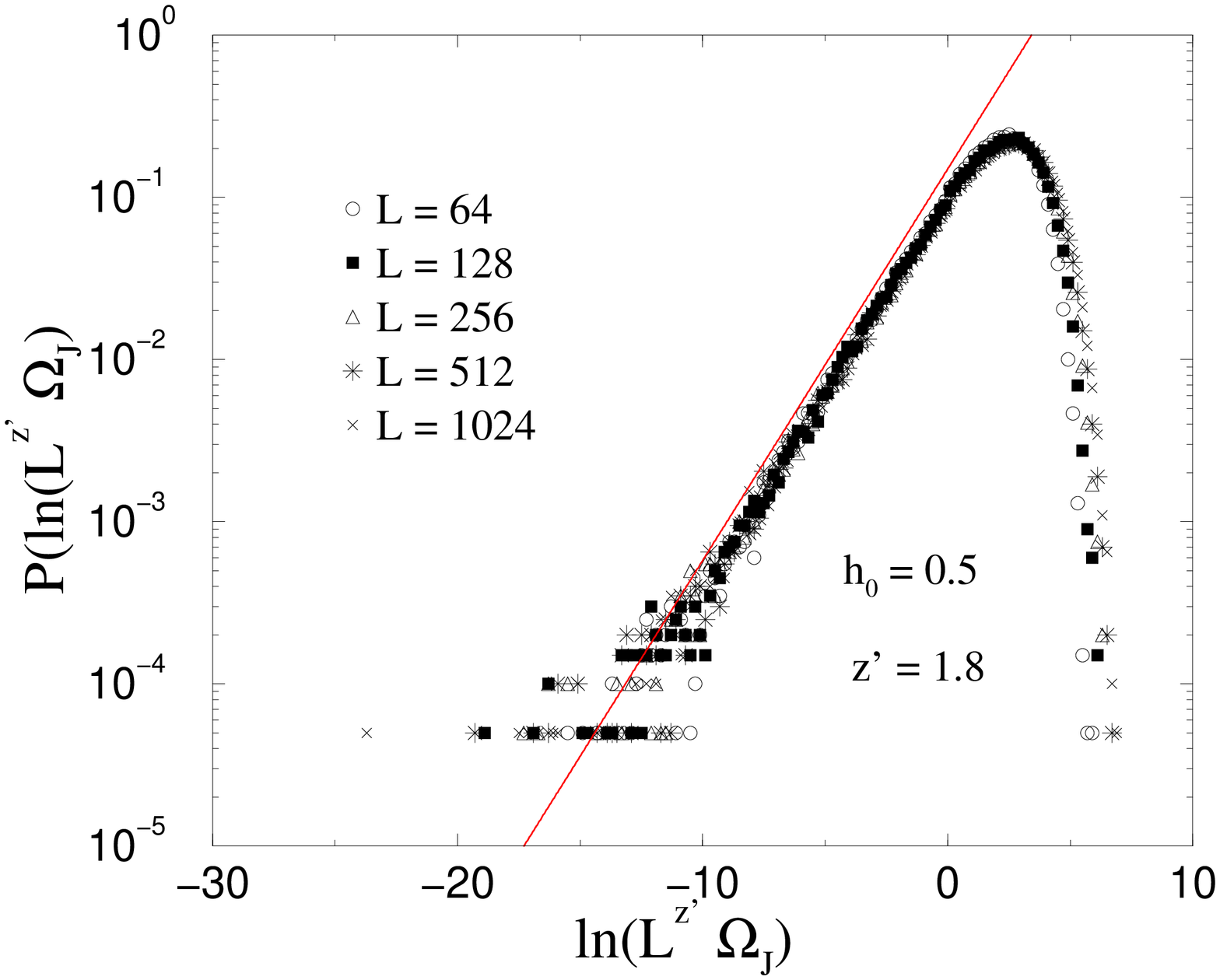} 
 \caption{Distribution of last log-bonds (left) in the ordered phase at
   $ h_{0}=0.5 $ for $d=1$ and its scaling plot (right). One observes that the
   scaling behavior of bonds in the ordered phase and that of fields in
   the disordered phase (see Fig.\ 5
   ) are related through duality.}
\end{figure}

\section{The double chain}

The RG scheme for double chains with some
new elements (compared to the 1d case treated
above) is depicted in Fig.\ 8.

\begin{figure}[!ht]
 \centering
 \label{dbs}
 \includegraphics[width=\halftext]{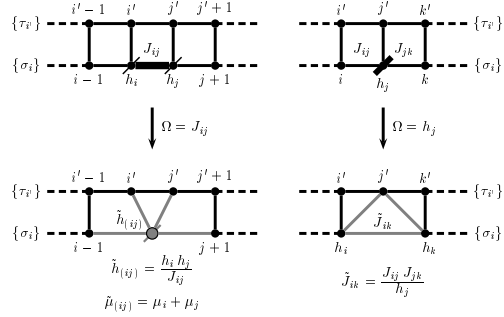}
  \caption{Schematic of renormalization-group decimation for double
    chains used in the numerical simulations.}
\end{figure}

As in $1d$, we observe that the log-field and log-bond distributions
get broader with increasing system size at criticality. To estimate
the critical point we compute the field distribution at the last stage
of the RG varying the initial transverse field $h_0$. We estimate the
critical point to be at $h_0=1.9$, beyond which the broadening of the
log-field distribution appears to be saturating, as for 1d in the
Griffiths phase. 
Moreover at $h_c=1.9$ the log-field and the log-bond
distributions become asymptotically identical
 except for a constant multiplicative factor
that reflects the short-ranged non-universal physics (See Fig.\ 9).
This is obvious in  the single chain, 
where it follows from the self-duality of the
simple chain at the critical point.  However, the double chain is {\it not} 
self-dual, nevertheless the scaling forms of the two distributions 
become identical at the critical point. We speculate that this remains
true also in the two-dimensional case to be discussed below.

\begin{figure}[!ht]
  \label{dbdis}
  \centering
  \includegraphics[angle=270,width=\halftext]{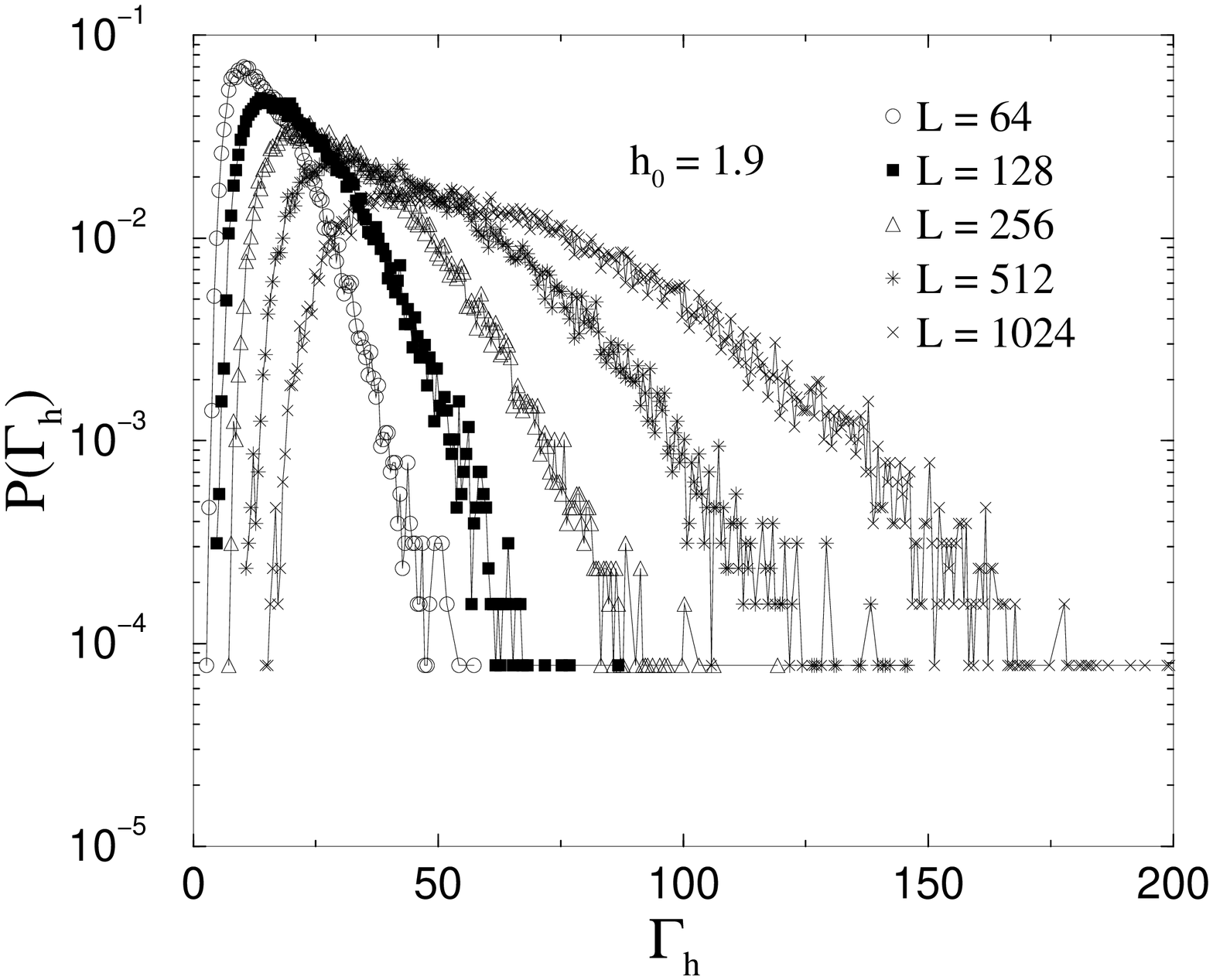}
  \hfill
  \includegraphics[angle=270,width=\halftext]{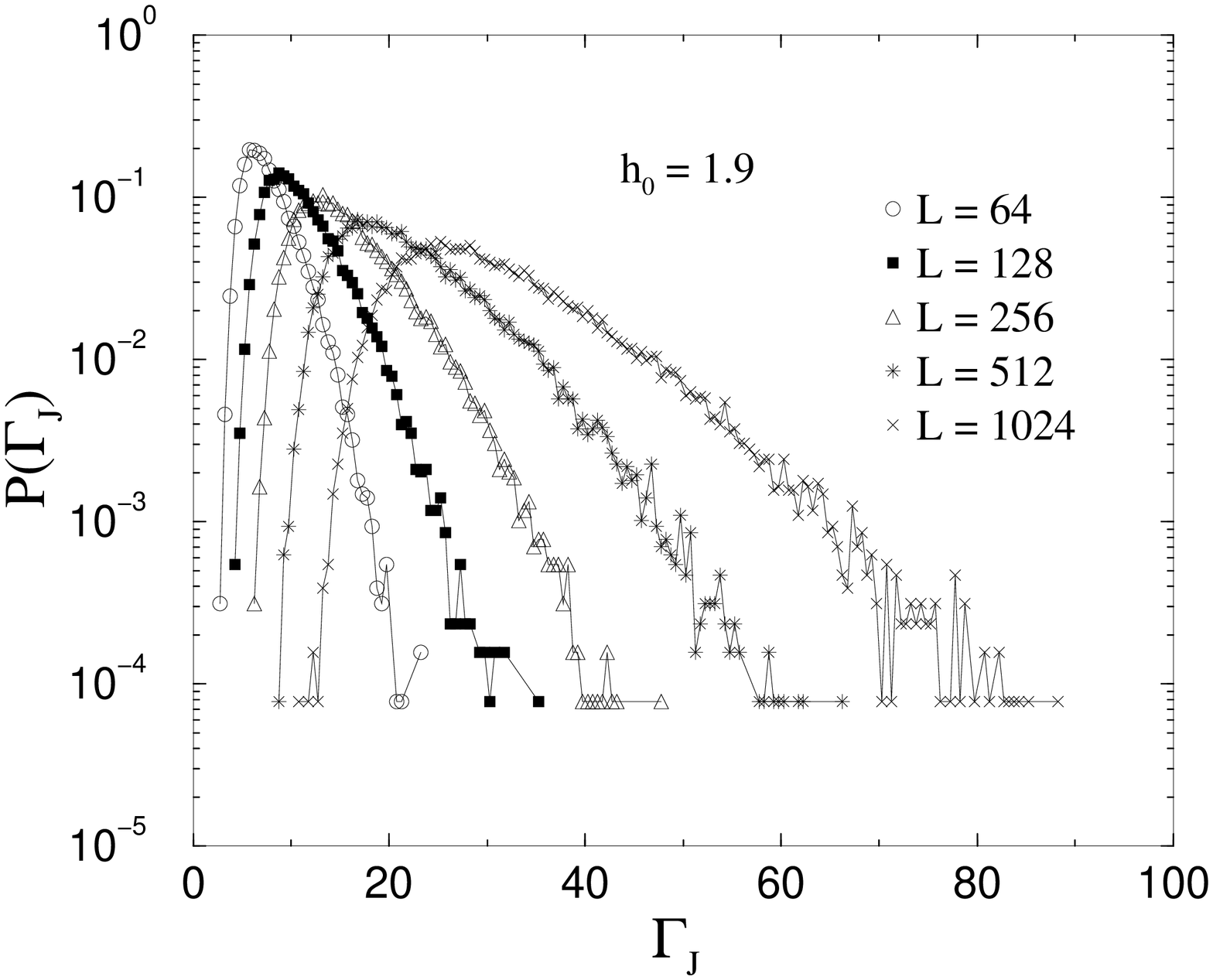} 
  \caption{
    The log-field (left) and log-bond (right) distribution for the
    double-chain at $h_c=1.9$. The data is obtained from $25\,600$ samples 
    for each system size.}
\end{figure}

\begin{figure}[!ht]
  \label{dbscale}
  \includegraphics[angle=270,width=\halftext]{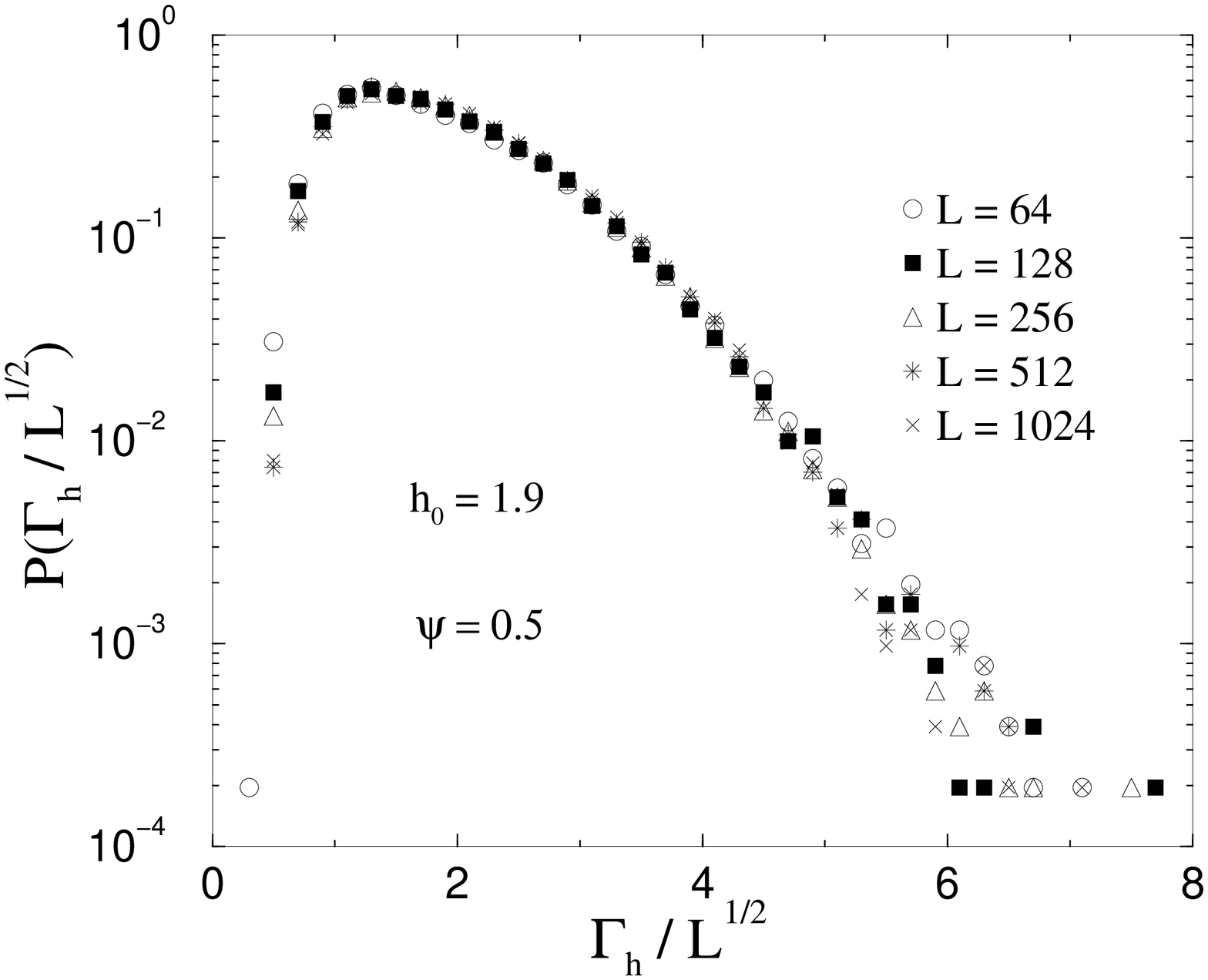}
  \hfill{}
  \includegraphics[angle=270,width=\halftext]{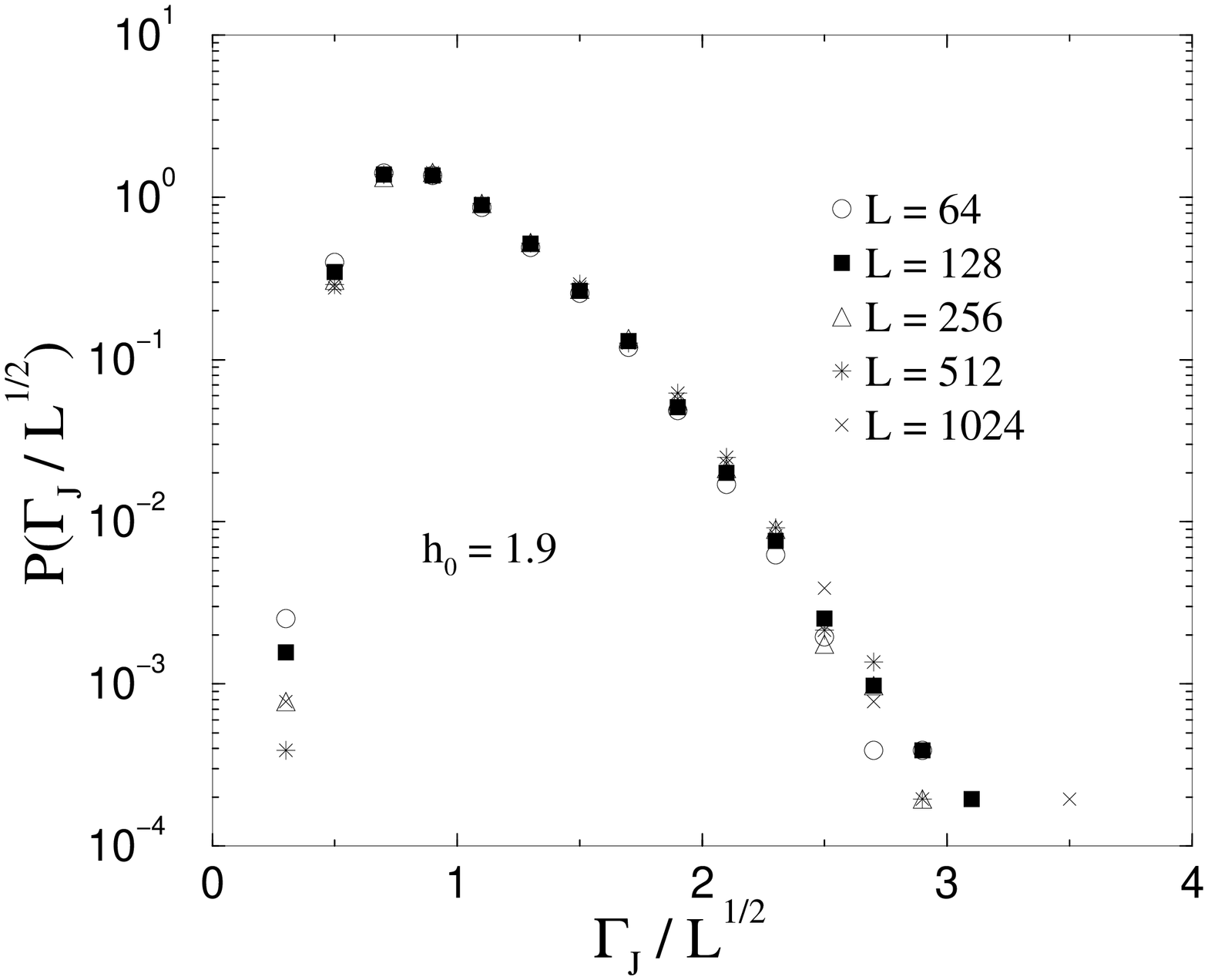}
  \caption{Scaling plots of the distribution of log-fields (left) and 
    of log-bonds (right) for the double-chain at the critical point
    ($ h_{c}=1.9 $). The data scales quite well with the same form
    $\Gamma _{h(J)}\equiv \ln (\frac{\Omega _{0}}{\Omega _{h(J)}})\sim
    \sqrt{L}$ as in 1d.}
\end{figure}

\begin{figure}[!ht]
 \label{dbm}
 \includegraphics[angle=270,width=\halftext]{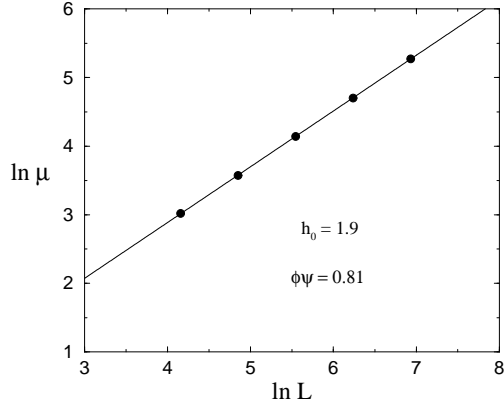}
 \begin{minipage}[t]{\halftext}
 \caption{A log-log scaling plot of the average magnetic moment per
   cluster, $\mu$, with the linear system size $L$ at the estimated
   critical point $h_c=1.9$ for the double chain. We find the same
   exponent $\phi$ as in $1d$.}
 \end{minipage}
\end{figure}

The scaling of the critical distributions depicted  in Fig.\ 9
yields the critical exponent $\psi=0.5$, as shown in Fig.\ 10.
This is the same as for the simple chain. In addition, for
the average magnetic moment of the last remaining cluster at $h_c=1.9$,
we find the same system size dependence for the double chain as for
the 1d case, i.e.\ the same critical exponent $\phi$, see Fig.\ 11. 
This implies that the double chain and the simple chain
belong to the same universality class.

\begin{figure}[!ht]
  \label{dbz}
  \includegraphics[angle=270,width=\halftext]{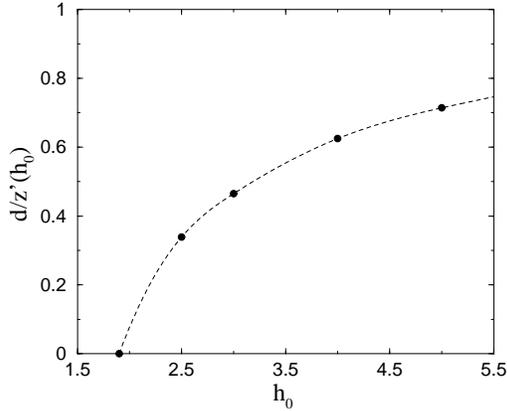}
  \begin{minipage}[t]{\halftext}
    \caption{Estimates for $1/z'$ against $h_0$ for the double chain.}
  \end{minipage}
\end{figure}

In the Griffiths phase $h_0>h_c=1.9$ we extracted the generalized
exponent $z'(h_0)$, which is depicted in Fig.\ 12.
Close to the critical point $h_c$ we observe the same linear
dependence of $1/z'(\delta)$ on the distance $\delta=h_0-h_c$ from the
critical point as in 1d. Since for $\delta\ll1$ one expects
$z'(h_0)\propto\delta^{-\psi\nu}$ this implies that $\nu=2$ the same
as the simple chain.

\section{The square lattice (2d)}

\begin{figure}[!ht]
 \centering
 \label{2ds}
 \includegraphics[width=\halftext]{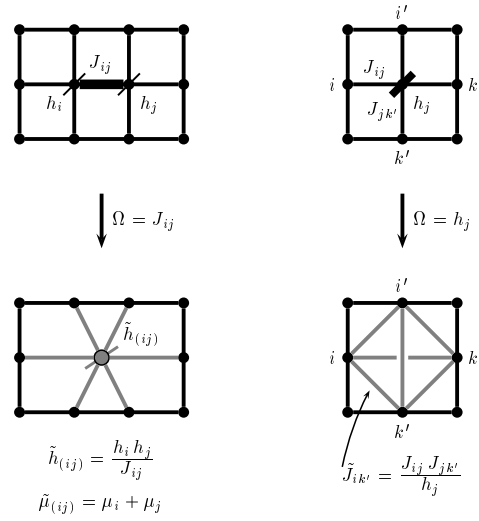}
 \caption{ Renormalization-group  decimations for the 
   two-dimensional (square) lattice used in the numerical simulations.}
\end{figure}

Next we present our preliminary results for the two-dimensional (2d)
case with periodic boundary conditions, where we, in contrast to the
treatment in Motrunich et al.\cite{motrunich}, keep {\it all} bonds
generated during renormalization.  The RG scheme for the
2d case is very similar to the one for the double chain
and is depicted in Fig.\ 13.

In comparison to 1d and the double chain models, the location of the
critical point cannot be fixed precisely for two dimensions according
to our numerical observation so far. We obtain a critical field
approximately at $h_0=5.3$ by applying the criterion that field and
bond distribution should have the similar scaling form (as for the 1d
case and the double chain). The scaling of the last log-field
distribution yields $\psi\approx0.5$ and the scaling plot of the
number of the active spins in the last remaining cluster yields $\phi
\approx 2.0$ and $\mu \sim L^{1.06}$.

\begin{figure}[!ht]
  \label{2da}
  \includegraphics[angle=270,width=\halftext]{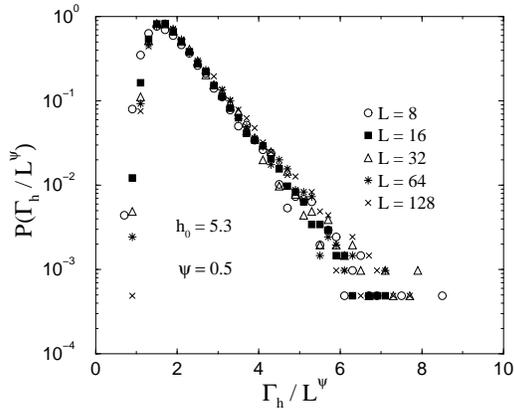}
  \begin{minipage}[t]{\halftext}
    \caption{Scaling of the log-fields for $d=2$ at the last stage of 
            the RG at what we estimate to be the critical point.}
  \end{minipage}
\end{figure}

\begin{figure}[!ht]
  \label{2db}
  \includegraphics[angle=270,width=\halftext]{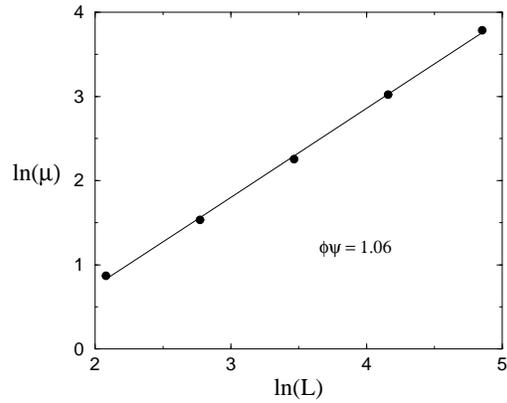}
  \begin{minipage}[t]{\halftext}
    \caption{A scaling plot of the number of the active spins in the last 
      remaining cluster at our candidate critical point $h_0=5.3$.}
  \end{minipage}
\end{figure}

Our preliminary results for the two-dimensional case agree with those
obtained recently by Motrunich et. al. \cite{motrunich} and with those
obtained by us via quantum Monte-Carlo simulation \cite{qmc}.

\section*{Acknowledgments}
H.R.\ is grateful to the German research foundation (DFG) for
financial support within a JSPS-DFG binational (Japan-Germany)
cooperation.
N.K.'s works is supported by Grant-in-Aid for Scientific Research 
Progam (No.11740232) from Mombusho, Japan.


\begin{thebibliography}{99}


\bibitem{mccoy}
  B. M. McCoy, \PRL{23,1969,383}.

\bibitem{fisher}
  D. S. Fisher,
  Physica A {\bf 263} (1999), 222.

\bibitem{motrunich}
  O. Motrunich, S.-C. Mau, D. A. Huse, D. S. Fisher, 
  cond-mat/9906322.

\bibitem{senthil}
  T.~Senthil and S.~Sachdev, 
  Phys. Rev. Lett {\bf 77} (1996), 5292.

\bibitem{youngrieger}
  A.P.~Young and H.~Rieger, \PR{B53,1996,8486}.

\bibitem{fisher1d}
  D.~S.~Fisher, Phys. Rev. Lett. {\bf 69} (1992), 534; 
  Phys. Rev. B {\bf 51} (1995), 6411.

\bibitem{big1d}
  F. Igl\'oi and H.\ Rieger,
  Phys. Rev. B {\bf 57} (1998), 11404.

\bibitem{ir2}
  F.~Igl\'{o}i and H.~Rieger, \PR{E58,1998,4238}.

\bibitem{oned}
  H. Rieger and F.\ Igl\'oi,
  Phys, Rev. Lett. {\bf 83} (1999), 3741;
  F.\ Igl\'oi, R. Juh\'asz and H.\ Rieger, 
  Phys. Rev. B {\bf 59} (1999), 11308;
  F.\ Igl\'oi, D.\ Karevski and H. Rieger,
  Europ.\ Phys.\ J.\ B {\bf 5} (1998), 613;
  H.\ Rieger and F.\ Igl\'oi,
  Europhys. Lett. {\bf 39} (1997), 135;
  F.\ Igl\'oi and H.\ Rieger,
   Phys. Rev. Lett. {\bf 78} (1997), 2473.

\bibitem{qmc}
  C. Pich, A. P. Young, H. Rieger, and N. Kawashima, Phys. Rev. Lett.
  {\bf 81} (1998), 5916; H. Rieger and N. Kawashima, Europ. Phys.
  J. B {\bf 9} (1999), 233; T. Ikegami, S. Miyashita and H. Rieger,
  J. Phys. Soc. Jap. {\bf 67} (1998), 2761.

\bibitem{dasgupta}
   S.\ K.\ Ma, C.\ Dasgupta and C.-K.\ Hu, Phys.\ Rev.\ Lett.\ 
  {\bf 43} (1979), 1434; 
  C. Dasgupta and S. K. Ma, \PR{B22,1980,1305}.

\end{thebibliography}
\end{document}